\documentclass[useAMS,usenatbib]{mn2e}

 \title[Ultracool dwarfs in UKIDSS]{The possiblity of detection of Ultracool
 Dwarfs with the UKIRT Infrared Deep Sky Survey}
 \author[N.R.\ Deacon \& N.C.\ Hambly]{N.R.\ Deacon\thanks{E-mail:
 nd@roe.ac.uk} and N.C.\ Hambly\\
SUPA\thanks{Scottish Universities Physics Alliance}, Institute for Astronomy, 
School of Physics, University of Edinburgh,\\
Royal Observatory Edinburgh, Blackford Hill,
 Edinburgh, EH9 3HJ\\}
 \begin{document}
 \date{}
 \pagerange{\pageref{firstpage}--\pageref{lastpage}} \pubyear{2005}
 \maketitle
 \label{firstpage}
 \begin{abstract}
 We present predictions for the numbers of ultra--cool dwarfs 
in the Galactic disk population
that could be
 detected by the WFCAM/UKIDSS Large Area Survey and Ultra Deep
 Survey. Simulated samples of objects are created with masses and ages drawn from different mass
 functions and birthrates. Each object is then given absolute magnitudes in
 different passbands based on empirically derived bolometric
 correction vs. effective temperature relationships (or model
 predictions for Y dwarfs). These are then combined with simulated
 space positions, velocities and photometric errors to yield observables such as
 apparent magnitudes and proper motions. Such observables are then passed through the
 survey selection mechanism to yield histograms in colour. This technique also produces predictions for the proper
 motion histograms for ultra--cool dwarfs and estimated numbers for the as yet
 undetected Y dwarfs. Finally it is shown that these techniques could
 be used to constrain the ultra low--mass mass function and birthrate
of the Galactic disk population. 
 \end{abstract}
 \section{Introduction} 
The precise nature of the physical mechanisms which result in the stellar
 Initial Mass Function (IMF) is one of the most fundamental open questions left
 in astrophysics at the start of this century. One of the main constraints that
 can be set on models of star formation is that the spectrum of masses (mass
 function) predicted is in good agreement with that observed in the local
 universe. The mass function $\xi(\log m)$
at any given time is defined to be the number of stars $dn$ per unit volume
in a mass interval $d\log m$. The definition of the mass
 function is, therefore, 
 \begin{equation}
 \label{MFdef1}
 \xi (\log m) = \frac{dn}{d \log m}.
 \end{equation}
Note that the mass {\em spectrum} 
defined by~\citet{Scalo} as $dn/dm$
 is often referred to as the mass {\em function} but we shall stick with
 the original definition in \citet{Salpeter}. In this paper, we will only
consider the IMF below $m\sim0.1$~M$_{\odot}$ as it is relatively poorly
constrained in this region. The two most common functional forms for the 
IMF are the power law form and the log--normal form. The power law form is
 \begin{equation}
 \xi \propto m^{-\alpha},
 \end{equation}
where above masses $m\sim1$~M$_\odot$, $\alpha\approx1.35$
 (e.g.~\citealt{Salpeter}) while at lower masses it is seen to flatten off.
 At the high mass end of our mass range, \citet{Chabrier2001}
 finds $\alpha\approx0.55$ while \citet{Kroupa} and \citet{Reid} derive
 $\alpha\approx0.3$. 
 Kroupa also fits a power law in the range $0.01 M_\odot < m <
 0.08 M_\odot$ finding $\alpha=-0.7$. Finally, \citet{Allen2005} use a
 series of assumptions about the birthrate and a Bayesian method to
 yeild a value of $-0.7$ in the range $0.04 M_\odot < m < 0.1 M_\odot$.
 In order to explore this range in power--law exponents, here
 we examine IMFs with values of $\alpha$ of~0, $-0.5$~and~$-1$. The
 log--normal form of the IMF is
 \begin{equation}
 \xi \propto exp \left( \frac{(\log_{10}m - \log_{10}m_c)^2}{2 \sigma ^2} \right).
 \end{equation}
 Here we use a log--normal function with the parameters $\log_{10}m_c=-1.1$ and
 $\sigma=0.79$ \citep{Chabrier2003}.

Regarding observational
studies of the IMF, those in open clusters benefit from all the stars in the
 cluster being of the same age. Hence, to derive the IMF
 it is simply a matter of converting a measured
 luminosity function (LF) using a 
 single mass--luminosity relation. However, objects in the field
 will have a range of ages, and since brown dwarfs
 (objects below the the hydrogen burning limit of $0.075M_\odot$) lack
 any internal energy source, they cool and decrease in brightness
 with time. In addition, the
 scale height of any stellar population within the Galactic disk
 evolves with time. Hence, both the photometric and kinematic
 properties of these objects are affected by their ages. In order to
 measure the mass function in the field we must therefore
 first consider the creation
 function and use models for the luminosity evolution of stars to
 convert it to an LF. This simulated LF can then be compared to the 
 observed LF to see if the creation function assumed is viable. 
 \citet{Miller} define the creation function $C$ as the number of objects created
 per unit time, per unit $\log m$ such that,
 \begin{equation}
 \label{CFdef}
 C(\log m,t) = \xi (\log m)\frac{b(t)}{T_{\rm G}},
 \end{equation}
 where $T_G$ is the age of the Galaxy and $b(t)$ is the
 stellar birthrate rate relative to the average birthrate
 such that $b(t)=(dn/dt)/(n_{\rm tot}/T_{\rm G})$, i.e.~the birthrate is the
 relative number of objects formed in the Galactic disk per unit time (note we assume that the IMF is 
 time--independent).
The \citet{Miller} study of the birthrate suggested that it does not
 depend strongly on the density of gas in the Galactic disk and is
 approximately constant. More recent studies such as
 \citet{RochaPinto} have shown that stars appear to form in a series of
 bursts. 
In this study we model the stellar birthrate as constant or as an exponential,
\begin{equation}
b(t) \propto e^{-t/\tau}.
\end{equation}
We employ four different values of the scale time $\tau$: three decreasing birthrates
 with $\tau=$ 10, 5 and 1 Gyr and one increasing with
 $\tau=-5$~Gyr. When generating age distributions in simulations we use
 10 Gyr as the maximum age for an object in the Galactic disk.

 In order to study the birthrate and the IMF below the hydrogen burning
 limit  a large sample of ultra--cool dwarfs is required. 
 Ultracool dwarfs are the observed L and T dwarfs and the as yet unobserved 
 Y~dwarfs. Infrared surveys are ideal for discovering large samples of cool 
 dwarfs as they are brightest in the near infrared.
The first major modern infrared surveys came with the Deep Near Infrared Survey
 (DENIS; \citealt{Epchtein}) and the Two Micron All Sky Survey
 (2MASS; \citealt{Kleinmann}). DENIS covers the southern sky in $I$, 
 $J$ and $K_s$ down to limits of $I=18.5$, $J=16.5$ and $K_s=14.0$ 
 while 2MASS is an all sky survey in $J$, $H$ and $K_s$ down to limits 
 of $J=15.8$, $H=15.1$ and $K_s=14.3$. Both
 have been useful for discovering ultra--cool dwarfs with DENIS finding several
 late M and L dwarfs and 2MASS finding countless L dwarfs and a large sample of
 T dwarfs. In addition several tens of L and T dwarfs~\citep{Chiu} have been found
 using the primarily optical Sloan Digital Sky Survey (SDSS, \citealt{Sloan}) 
The next generation of infrared surveys will be undertaken with large
format imagers such as 
WFCAM~\citep{Henry} 
and WIRCAM~\citep{Puget}. WFCAM is a wide--field quasi Schmidt camera 
mounted at the Cassegrain focus of the
 UK Infrared Telescope (UKIRT). A quadruple detector array provides (with
 mosaiced observations) a field of view of 0.77 sq. degrees. 
The UKIRT Infrared Deep Sky Survey (UKIDSS; \citealt{UKIDSS}) 
is employing WFCAM for a series of Galactic and
 extragalactic surveys, and two of these surveys are ideal for the detection of
 ultra--cool dwarfs in the field. The Large Area Survey (LAS) is a
 wide--field, high latitude survey in $Y$~\citep{Hewett} and
$JHK$ in the MKO system (\citealt{Tokunaga}; \citealt{Simons}); the Ultra Deep
 Survey (UDS) is a deep, narrow--field survey in $JHK$. 
 The initial two year program for the LAS will be 2000
 square degrees to a depth of $J=19.7$. The seven year plan will
 include a second $J$ band epoch to allow proper motion measurements and
 will cover 4000 square degrees to a depth of $J=20$. The UDS will
 cover 0.77 sq. degrees in both the two and seven year plans. In the two year plan
 only $J$ and $K$ will be observed to a depth of $J=24.0$,
 while the full seven year plan also includes $H$ and will go one magnitude 
 fainter in $J$. The number of objects of different effective temperatures and
 colours (which we shall employ as a detection function) found in these 
surveys can provide information on the IMF (e.g.~\citealt{Pinfield}) 
and birthrate.

As well as having important statistical properties, ultra--cool dwarfs
 are also interesting objects in themselves. One particular area that
 UKIDSS hopes to study is the transition between the L and T spectral
 types. Here the dramatic onset of methane absorption
around $2\mu m$ and the removal of dust
 clouds high in the photosphere cause a sudden shift from
 red near infrared colours to blue near infrared colours. The exact details of
 this transition are still a matter for debate with many theories such
 as patchy cloud clearing \citep{Burgasser2002}, sudden downpours 
\citep{Knapp} and runaway dust growth \citep{Tsuji} seeking to explain it.
Also of interest are the as yet unobserved Y dwarfs. It is not yet
 known where there will be a significant shift in atmospheric
 chemistry causing the use of the label Y. Hence we define Y dwarfs as
 cooler than the temperature for the coolest subclass of T dwarf
 given by~\cite{Vrba}, 770K. There are many
 different sets of models -- e.g.~\citet{Burrows}, 
 \citet{Baraffe}, \citet{Marley} and \citet{Tsuji} --
 which predict a range of near infrared colours for these objects~\citep{Leggett}. 
Clearly Y dwarfs discovered with UKIDSS can help to constrain
 these atmosphere models. 
 
 In the rest of this paper, we go on to describe our simulation method 
 for predicting
 the number of ultra--cool dwarfs that will be found in the UKIDSS~LAS
 and~UDS given the various assumptions described above concerning the 
 form of the underlying birthrate and IMF.

 \section{Simulation Method}
\label{sim}

 We implement a simulated population method to predict the possible
 results of the UKIDSS LAS and UDS. Simulated 
 (very low mass) stellar and substellar
 populations based on the different IMFs and birthrates discussed previously are
 created. 

 Each simulated object is given an age based on the birthrate and a mass based on the IMF,
 and these dictate the photometric and astrometric properties assigned.
 Each object is subsequently passed through the survey selection mechanism to yield
 the simulated results of the survey. The local number of very low mass stars 
 simulated is anchored using the same method as \citet{Burgasser} with 
 the mass spectrum in
 the range $0.1-0.09$~M$_\odot$ set to $0.0055$~pc$^{-3}$. Using this method,
 histograms binned by observables such as proper motions and colours can be
 produced.
 
 \subsection{Positional and Velocity Simulations}
 
 In order to effectively model the positions and proper motions of our synthetic
 stellar population we must first consider their cartesian positions and space
 velocities. We generate heliocentric Galactic cartesian coordinates out to a
 maximum distance of 2~kpc for the LAS and 10~kpc for the UDS.
 Our simulations indicate that L dwarfs are not detectable in either
 survey at distances greater than these. In our coordinate system $x$ and $y$
 lie in the Galactic plane while $z$ is perpendicular to the plane. These have
 velocities $U$, $V$ and $W$ associated with them. The $x$ and $y$ positions are
 randomly generated from a flat distribution while the $z$ position is generated
 from an exponential distribution with scale height $z_0$. It is well established
 that the older populations have larger scale heights, hence we must incorporate
 this age dependency into our model. It is also clear that due to disk heating a
 population's velocity dispersion increases with age, and this is taken into
 account. Hence, we generate space positions and velocities based on the age of
 the object. From these we calculate the sky positions, distance and proper
 motions. Any object whose angular position falls within the survey area
 passes the positional survey selection mechanism.

 \subsection{Photometric Simulation}
 
\begin{figure}
 \setlength{\unitlength}{1mm}
 \begin{picture}(75,150)
 \includegraphics{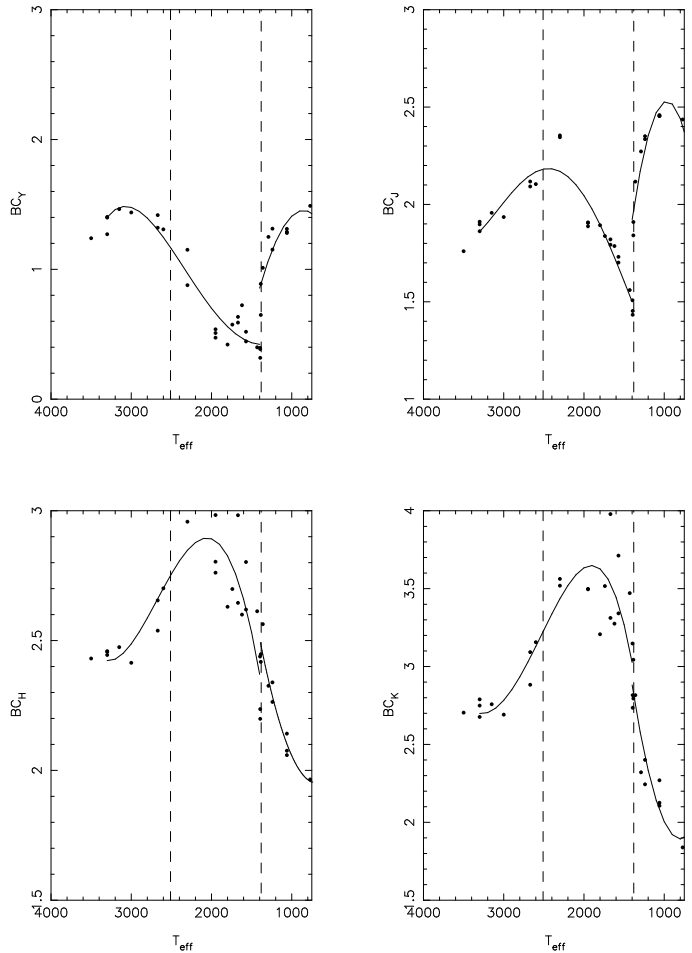}
 \end{picture}
 \caption[]{The bolometric correction-effective temperature relation
 for the Y \citep{Hewett} and JHK (MKO) passbands (see text).}
 \label{BC}
 \end{figure}

 We utilise evolutionary models from~\citet{Baraffe} of ultra--cool dwarfs to provide look-up tables
 of quantities such as~$T_{\rm eff}$ and absolute magnitudes in a given
 passband, versus mass and age.
 The current picture of substellar models is one of division between two
 different model sets: those in which dust is suspended in the
 photosphere and those where the dust has largely settled below the photosphere. 
 The spectrum of T dwarfs is best fitted by the models where the dust has settled,
 but the infrared colours of such models do not match well those observed for L dwarfs. 
 We use temperatures and bolometric magnitudes from the dust settled (COND)
 models~\footnote{Solar metalicity only.}
 across the M, L, T and Y regimes in
 a manner similar to \citet{Burgasser} (who noted that this would not
 produce a temperature error greater than $10\%$). However, in order
 to properly model the photometric detectability
 for all these spectral types we must utilise an effective temperature versus bolometric 
 correction relation of reach passband. \citet{Hewett} calculated the colours of a series of 
 ultra--cool objects from their spectra. We combined these with absolute J 
 magnitudes (converted into the MKO system) along with absolute
 magnitudes and effective temperatures~\footnote{We ignore any
 uncertainty in the effective temperature values due to the unknown ages of
 some of the objects they are based on.} from \citet{Vrba}
 and bolometric magnitudes from \citet{Golimowski}. We then
 fitted two separate polynomials in the T regime and in the M \& L regime
 (Figure~\ref{BC}) since it is clear that in some passbands the
 relationship is discontinuous across the L--T boundary. We utilise the
 \citet{Baraffe} models in which the dust has settled below the
 photosphere to provide an effective temperature and a bolometric magnitude
 for each of the simulated stars. The temperature is then used to
 calculate the bolometric corrections (and hence infrared magnitudes) for each
 simulated object. For objects cooler than $T_{\textrm{eff}}=770$ K we have no
 observational data on which to base bolometric corrections so we rely on the
 magnitudes predicted by the models. Finally,
 the photometric survey selection method is simply detection above the
 quoted depths for the survey in all passbands used except K. This is
 because cool T dwarfs will be very faint in the K band due to
 methane and water absorbtion. 

 \subsection{Simulated \boldmath$T_{\rm eff}$ Distributions}
 
\begin{figure}
 \setlength{\unitlength}{1mm}
 \begin{picture}(75,75)
 \includegraphics{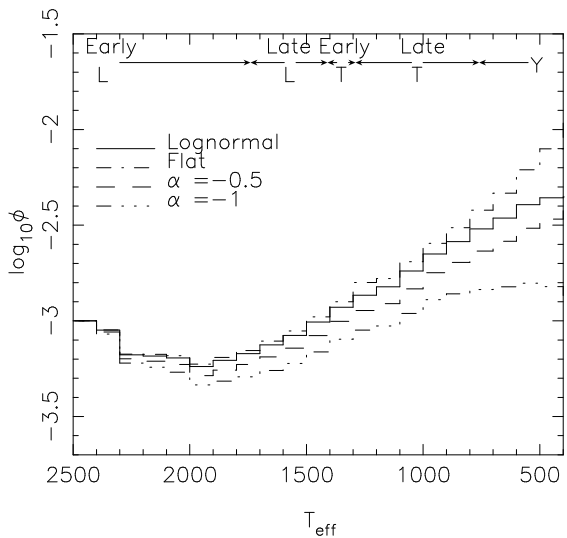}
 \end{picture}
 \caption[]{The alteration to the effective temperature distribution caused by 
 different underlying IMFs. In this case a constant birthrate is used
 and the simulations are all normalised to the same point in the
 hottest bin. 
 Note that the main effect is the height of the second peak. The
 temperature regions corresponding to each spectral type are taken
 from~\citet{Vrba}.}
 \label{MFchange}
 \end{figure}

\begin{figure}
 \setlength{\unitlength}{1mm}
 \begin{picture}(75,75)
 \includegraphics{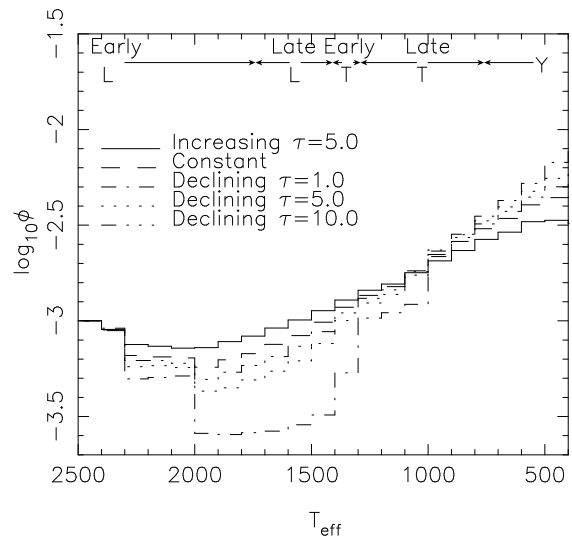}
 \end{picture}
 \caption[]{The alteration to the effective temperature distribution caused by 
 different underlying birthrates. In this case a log--normal IMF is
 used and again the simulations are all normalised to the same point in the
 hottest bin. 
 Note that the main effect is the depth of the trough. Again the
 temperature regions corresponding to each spectral type are taken
 from~\citet{Vrba}.}
 \label{birthchange}
 \end{figure}

 The simulated $T_{\rm eff}$ distributions shown in Figures~\ref{MFchange} 
 and~\ref{birthchange} illustrate how altering the IMF and 
 birthrate respectively affects such histograms. Mass functions which increase at lower masses will
 increase the number of low luminosity objects and hence will raise
 the height of the peak around  $500$~K. A birthrate where most
 objects were created at early star formation epochs will deepen the trough
 around $1700$~K as most objects will have evolved past the mid
 L spectral types into the T and Y domain. These peaks and troughs are
 similar to those found by \citet{Burgasser} but, due to the varying
 ages of the objects, the complicated system of peaks and troughs
 found by \citet{Allen} for single--age populations is not seen. 
 Since the underlying form of the IMF and birthrate affects markedly
 the number counts in $T_{\rm eff}$ and colour, those histograms
 should be a useful probe of the IMF and birthrate.

  
 \section{The Large Area Survey}
 
\begin{figure}

 \setlength{\unitlength}{1mm}
 \begin{picture}(75,130)
 \includegraphics{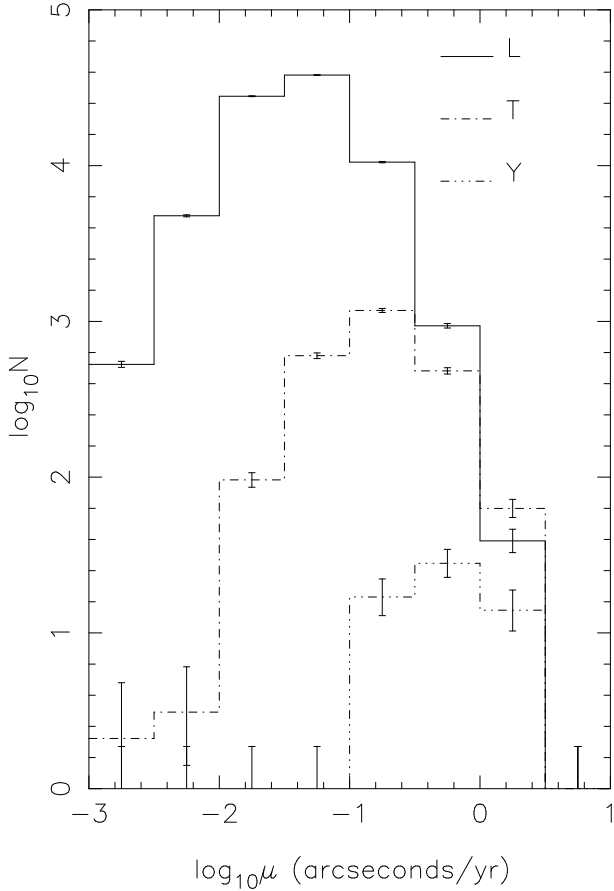}
 \end{picture}
 \caption[]{Proper motion histograms for objects of different spectral
 types in the seven year ($Y=20.5$, $J=20.0$, $H=18.8$, $K=18.4$ over
 4000sq. degrees) UKIDSS LAS assuming a log--normal IMF and a constant birthrate.}
 \label{PMhist}
 \end{figure}

 The LAS is designed to search for ultra--cool dwarfs, high redshift
 quasars and cool subdwarfs. It uses $J$, $H$ and $K$ in addition to the $Y$
 filter~\citep{Hewett}. The $Y$ filter lies in between the $I$ and $J$ bands, is centred around
 one micron and is slightly redder than the $Z$ filter. It is specifically designed
 for the study of ultra--cool dwarfs and quasars. It will provide a
 band to allow M, L, T and Y dwarfs to be distinguished both from each
 other and from hotter main sequence stars. Additionally
 SDSS~\citep{Sloan} photometry can be used to remove quasars from the sample~\citep{Hewett}. The final survey will cover an area
 around the northern Galactic pole as well as a small strip in the
 south Galactic cap. Both areas are scanned by the Sloan Digital Sky
 Survey (SDSS, \citealt{Sloan}) 
 allowing additional optical photometry to be utilised. The full seven
 year survey will go
 to depths of $Y=20.5$, $J=20.0$, $H=18.8$ and $K=18.4$ (detection in
 $K$ was not required in the survey selection mechanism as many cooler
 objects have their luminosity severly reduced in this band due to methane and
 water absorption) with an additional
 second scan in $J$ to allow proper motion measurements.
 Simulations of the full seven year LAS area were carried out as described in
 Section~\ref{sim}. Figure~\ref{PMhist} shows the simulated proper
 motion histograms for objects of different spectral types. Notice
 that later spectral types' histograms peak at higher proper
 motions. This is simply a selection effect as cooler objects will
 only be observable nearby, where they will typically have large
 proper motions\footnote{Here (as in all the LAS simulations) we
 require detection in $Y$, $J$ and $H$.}.

\begin{figure}
 \setlength{\unitlength}{1mm}
 \begin{picture}(75,130)
 \includegraphics{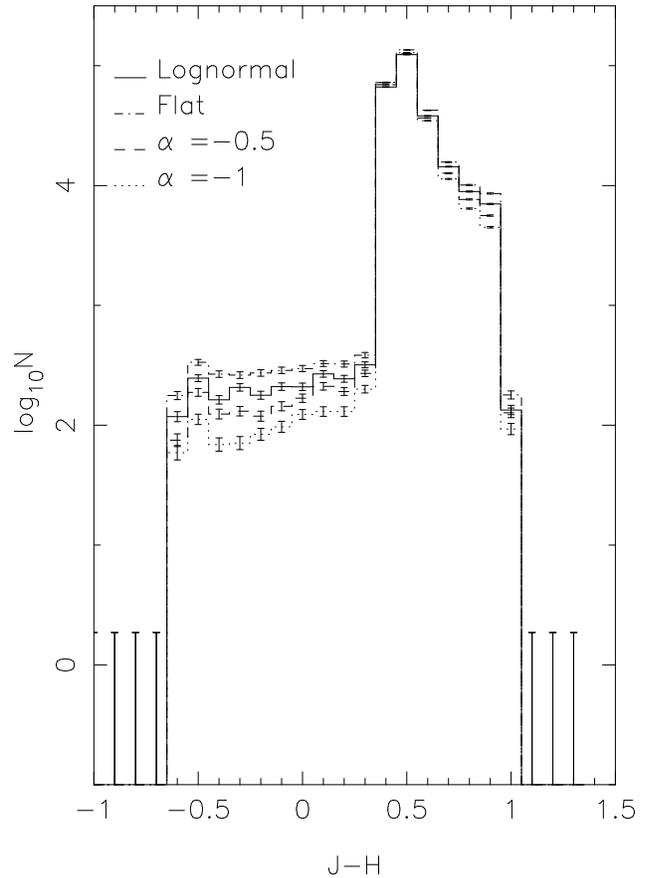}
 \end{picture}
 \caption[]{The alteration of the colour distribution caused by
 different Mass Functions. A constant birthrate is used here. Note the
 large step around $(J-H)=0.4$. This marks the boundary between the
 Ys and mid-late Ts (to the left) and the L and early Ts (to the
 right). All J,
 H, and K magnitudes are in the MKO system (\citealt{Tokunaga}, \citealt{Simons}).}
 \label{MFalter}
 \end{figure}

 The expected numbers of detected objects of different spectral types 
 are shown in Table~\ref{LASnumb}.
\begin{table*}
 \centering
 \begin{minipage}{140mm}
  \caption{The number of objects of different spectral types for
 varying birthrates and IMFs for the seven year UKIDSS LAS. Note that the
 $\tau=1 \rm Gyr$ birthrate is included to illustrate the effect of a
 changing scale time. We do not consider it to be a realistic distribution.}
\label{LASnumb}
  \begin{tabular}{lllllll}
  \hline
Mass & Birthrate & Early L & Late L & Early T & Late T &Y\\
Function&&Dwarfs&Dwarfs&Dwarfs&Dwarfs&Dwarfs\\
  \hline
Log--normal&Constant&77484&5620&968&1444&59\\
$\alpha=0$&Constant&85182&6892&1235&2014&100\\
$\alpha=-0.5$&Constant&72476&4420&752&1060&39\\
$\alpha=-1$&Constant&67947&3479&513&676&25\\
  \hline
Log--normal&$\tau=-5.0$&53559&8648&1375&1893&57\\
Log--normal&Constant&77484&5620&968&1444&59\\
Log--normal&$\tau=10.0$&86296&4955&860&1276&42\\
Log--normal&$\tau=5.0$&93316&4096&778&1173&46\\
Log--normal&$\tau=1.0$&120515&1517&321&657&31\\
  \hline
\end{tabular}
\end{minipage}
\end{table*}
Figure~\ref{MFalter} shows the effect on the $(J-H)_{MKO}$ colour\footnote{We use
 the $J-H$ colour as we did not require a detection in $K$ and the
 predicted $Y-J$ magnitudes for T dwarfs are all very similar.} histogram of altering
 the IMF. The sharp drop at $(J-H)_{MKO}=0.4$ is due to objects
 bluer than this divide being mid T dwarfs and objects
 redder than this being (on the whole) the much more easily detectable early
 L dwarfs. It is clear that for more steeply declining IMFs fewer
 T dwarfs are observed. Note that the log--normal and flat ($\alpha=0$) IMFs
 produce very similar results.

\begin{figure}
 \setlength{\unitlength}{1mm}
 \begin{picture}(75,130)
 \includegraphics{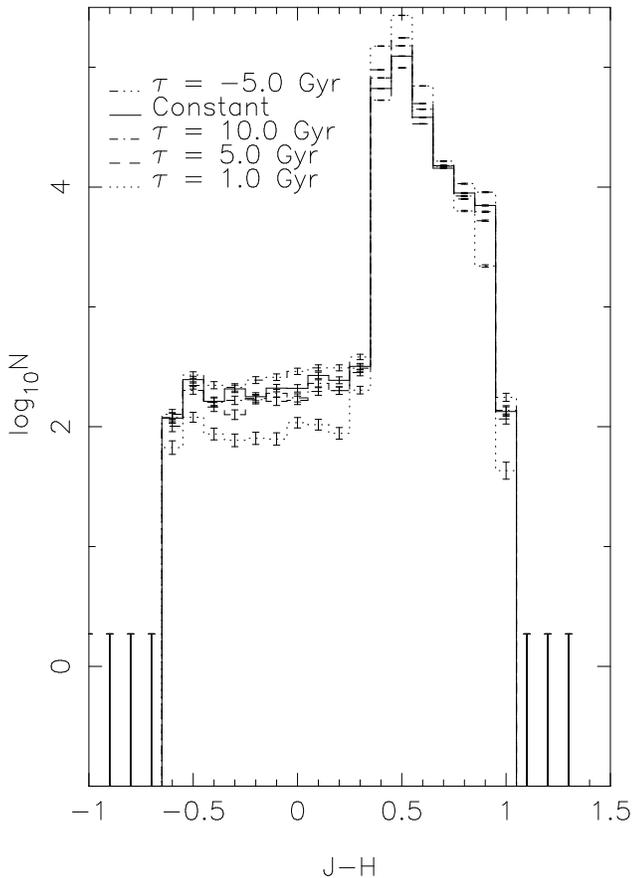}
 \end{picture}
 \caption[]{The alteration of the colour distribution caused by
 different birthrates. A log--normal Mass Function is used here. Note the
 large step around $(J-H)=0.4$. This marks the boundary between the
 Ys and mid-late Ts (to the left) and the L and early Ts (to the
 right). J and~H
 magnitudes are on the MKO system (\citealt{Tokunaga}; \citealt{Simons}).}
 \label{balter}
 \end{figure}

 The effects of different birthrates are shown in
 Figure~\ref{balter}. Here, birthrates which produced more objects
 earlier in the Galaxy's history have fewer T dwarfs due to the
 cooling of brown dwarfs with time. The higher
 numbers of early L dwarfs in simulations with birthrates which were
 higher in the past is due to our normalisation and the way we have
 modelled scale height evolution. As the age of a population increases
 it becomes more spread out and hence the density in the Galactic
 plane drops. Since we are normalising in the local region (the Galactic
 plane) the number of objects here is kept constant. Hence as the
 population spreads the total number of objects increases. So if a
 particular class of object has enough mass for stable hydrogen
 burning, as early L dwarfs do, the number of detectable objects is
 increased by a birthrate which was historically higher. Distributions
 with very long scale times (either increasing or decreasing) 
 would be difficult to distinguish from a constant birthrate. Note
 that in all cases the numbers of early T dwarfs are much smaller than
 those for late T dwarfs. This appears counterintuitive as early T
 dwarfs are brighter and hence more detectable. However, as the
 atmospheric chemistry of early T dwarfs changes quickly with
 effective temperature the spectral types T0-T4 cover a very small
 region of an effective temperature distribution (barely 100K, see
 Figures~\ref{MFchange} and~\ref{birthchange}). Hence even though late
 T dwarfs are more difficult to detect than early T dwarfs, their
 larger temperature spread (and hence higher numbers) lead to a
 greater number of detections. 

 \section{The Ultra Deep Survey}
\label{UDS}

The UDS is a deep pencilbeam survey, covering 0.77 sq. degrees, primarily designed for
extragalactic studies. 
For this purpose it uses the $J$, $H$ and $K$ passbands down to depths of
 $J=25$, $H =24$ and $K=23$. This presents a problem when trying to
 detect ultra--cool dwarfs. T and Y dwarfs have similar near infrared colours to
 main sequence stars due to methane absorption around two microns. Hence with
 only $J$, $H$ and $K$ photometry detection can be difficult. Luckily
 for the study of ultra--cool dwarfs the area
 covered by the UDS is also covered by the Subaru/XMM-Newton Deep Survey (SXDS)
 which provides optical photometry. This will make it easier to distinguish
 ultra--cool dwarfs from other objects. Of course, many of the individual
 ultra--cool dwarfs found in the UDS will be too faint for
 spectroscopic follow-up observations. However they will still contribute to the
 observed sample. Note that
this survey is so deep that the Galactic disk scale length had to be taken
into account along with the scale height; 
we used a value of $3.5 \rm kpc$~\citep{Vaucouleurs}.
The results are shown in Table~\ref{numbUDS},
where we see that a few tens of T dwarfs, a few hundred L dwarfs 
and a handful of Y dwarfs will be detected. 
While this sample will not be as useful as that of
the LAS for studying the IMF, birthrate and spectroscopic
properties of these objects, it may provide valuable data on their
distribution within the Galaxy. 

\begin{table*}
 \centering
 \begin{minipage}{140mm}
  \caption{Predicted numbers of objects of different spectral types, for
 various birthrates and IMFs, in the seven year UKIDSS UDS.}
\label{numbUDS}
  \begin{tabular}{lllllll}
  \hline
Mass & Birthrate & Early L & Late L & Early T & Late T &Y\\
Function&&Dwarfs&Dwarfs&Dwarfs&Dwarfs&Dwarfs\\
  \hline
Log--normal&Constant&129&72&26&102&21\\
$\alpha=0$&Constant&228&135&42&208&58\\
$\alpha=-0.5$&Constant&72&34&14&47&7\\
$\alpha=-1$&Constant&85&43&12&45&4\\
  \hline
Log--normal&$\tau=-5.0$&103&69&23&106&15\\
Log--normal&Constant&129&72&26&102&21\\
Log--normal&$\tau=10.0$&250&113&37&129&23\\
Log--normal&$\tau=5.0$&288&103&39&125&23\\
Log--normal&$\tau=1.0$&465&59&24&66&3\\
  \hline
\end{tabular}
\end{minipage}
\end{table*}
\section{Constraining the IMF and Birthrate}

In the previous two Sections we have shown that it is possible to
predict the numbers of objects detected in UKIDSS surveys using
different birthrates and IMFs. Clearly, we can 
attempt also the reverse, and use the observed numbers to constrain the
underlying birthrate and IMF. For example, suppose that we
assume that the IMF (in the range
$0.1-0.003 M_\odot$) and birthrate have the following functional
forms:
\begin{equation} 
\begin{array}{c}
b(t) \propto e^{- \beta t},\\
\xi \propto m^{- \alpha},
\end{array}
\end{equation}
where a positive value of $\beta$ implies a declining birthrate. We
now simulate a grid of $J-H$ distributions with values of $\alpha$
ranging from --2.0 to +2.0 and $\beta$ ranging from --0.2 to
+0.2. We then take another simulated distribution with known $\alpha$
and $\beta$. A multiplication factor $\gamma$ is then used to vary
the total number of objects for each simulated 
$(J-H)$ distribution. The values of $\chi^2$ for a range of
values of $\gamma$ are calculated for each value of $\alpha$ and
$\beta$. Hence we get a value of $\chi^2$ for each value of $\gamma$
and for each $(J-H)$ histogram (and hence for each value of $\alpha$ and
$\beta$). This distribution is then marginalised over $\gamma$
(i.e. the probability distribution is integrated over $\gamma$) such
that
\begin{equation}
p(\alpha,\beta)= \frac{\int e^{-\chi^2/2}d\gamma}{\int \int \int
  e^{-\chi^2/2}d\gamma d\alpha d \beta}.
\end{equation}
This procedure produces a probability distribution over $\alpha$
and $\beta$. Marginalising this further will produce distributions
solely  over $\alpha$ or $\beta$.  

\begin{figure}
 \setlength{\unitlength}{1mm}
 \begin{picture}(75,100)
 \includegraphics{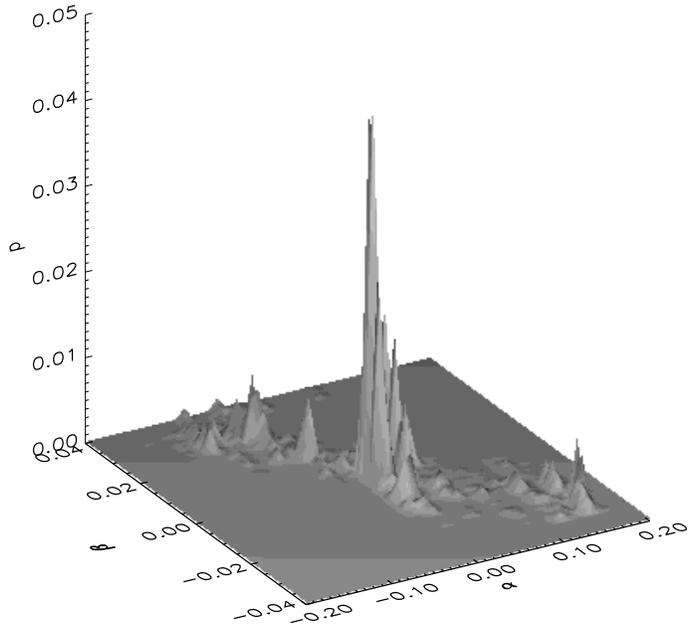}
 \end{picture}
 \caption[]{The probability surface produced by the comparing an
 $\alpha=0$, $\beta=0$ distribution to a finer grid around the peak of
 the coarse grid. There is clearly a degeneracy between $\alpha$ and
 $\beta$.}
 \label{zerozoom}
 \end{figure}

\begin{figure}
 \setlength{\unitlength}{1mm}
 \begin{picture}(75,100)
 \includegraphics{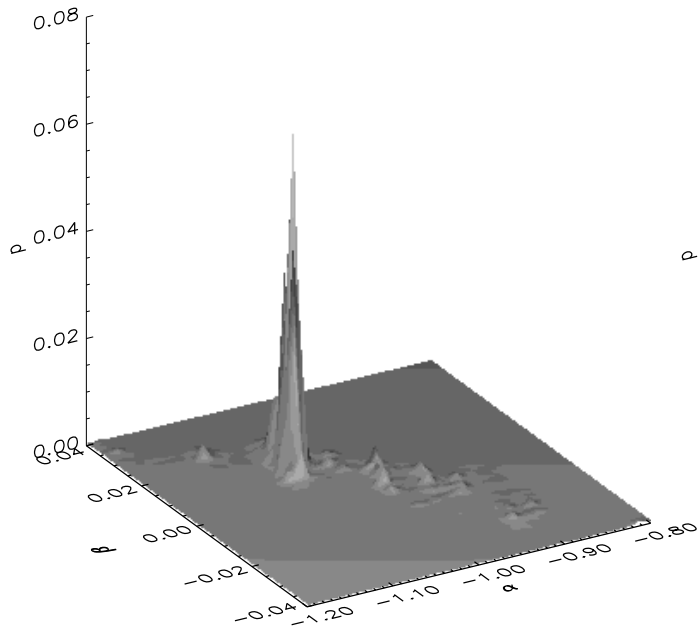}
 \end{picture}
 \caption[]{The probability surface produced by the comparing an
 $\alpha=-1$, $\beta=0$ distribution to a finer grid around the peak of
 the coarse grid. The degeneracy between $\alpha$ and
 $\beta$ can be seen.}
 \label{minus1zoom}
 \end{figure}

\begin{figure}
 \setlength{\unitlength}{1mm}
 \begin{picture}(75,100)
 \includegraphics{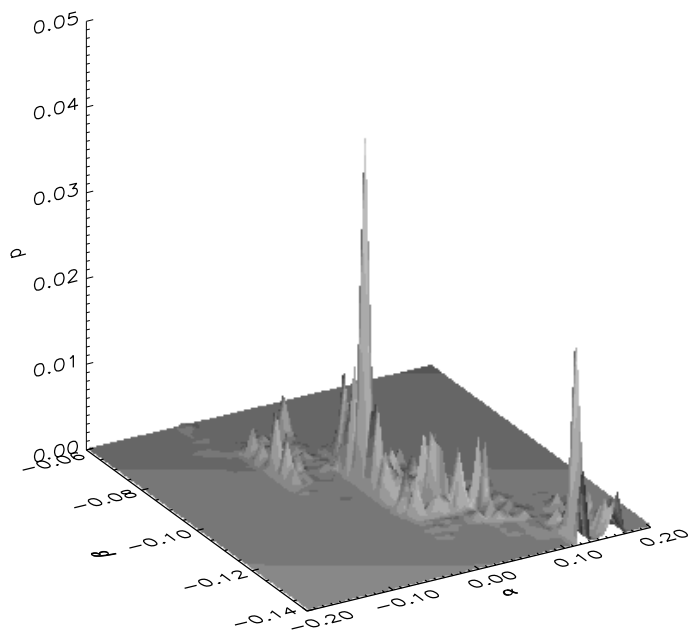}
 \end{picture}
 \caption[]{The probability surface produced by the comparing an
 $\alpha=0$, $\beta=-0.1$ distribution to a finer grid around the peak of
 the coarse grid. Again the degeneracy between $\alpha$ and
 $\beta$ is clear.}
 \label{0.1zoom}
 \end{figure}

\begin{table*}
 \begin{minipage}{140mm}
\centering
  \caption{The calculated values for $\alpha$ and $\beta$ for a range
  of different input values.}
\label{res1}
  \begin{tabular}{cccc}
  \hline
Input& Input&Calculated& Calculated\\
$\alpha$&$\beta$&$\alpha$&$\beta$\\
  \hline
0&0&0.021 $\pm$ 0.060&0.003 $\pm$ 0.015\\
-1&0&-1.038$\pm$0.042&0.012$\pm$0.009\\
0&-0.1&0.031$\pm$0.056&-0.094$\pm$0.017\\
  \hline
\end{tabular}
\end{minipage}

\end{table*}
In order to study the constraints that could be set in more detail we
first simulate a coarse grid in the ranges $-2.0<\alpha<2.0$ and
$-0.2<\beta<0.2$. Once we identify the area of maximum probability in
this coarse grid we simulate a second, finer grid centred around this region. This has ten times
the resolution in $\alpha$ and five times the resolution in
$\beta$. The results for such finer grids are shown in
Figure~\ref{zerozoom} ($\alpha =0, \beta=0$), Figure~\ref{minus1zoom}
($\alpha= -1, \beta=0$) and Figure~\ref{0.1zoom} ($\alpha=0,
\beta=-0.1$). The first thing that becomes apparent is that the noise on these
finer grids means that there is not a smooth probability
distribution. However it is obvious that there is a degeneracy between
$\alpha$ and $\beta$. In order to glean information on the typical
errors expected the probability distribution was marginalised over
$\alpha$ to produce a distribution in $\beta$ and vice-versa. The mean
value of each parameter along with their standard deviations can then
be calculated. These results are shown in Table~\ref{res1}. It is
clear that the calculated values are in agreement with the values of
the parameters used to generate the $J-H$ histograms.
\subsection{Constraining the IMF and birthrate using existing data}

To allow some real results in advance of UKIDSS data
being available, we simulated the
results of an existing ultra--cool dwarf survey, {\em viz.}~the
$J$ band luminosity function of \citet{Cruz2003}, covering the L dwarf
regime. We
simulated a grid in the same manner as for the UKIDSS LAS with the
appropriate cuts and limiting magnitudes quoted in Cruz et al. We
excluded the bin centred on $M_J = 10.75$ as, although our maximum
mass of $0.1M_\odot$ equates to a stable main sequence absolute $J$
magnitude of 10.2~\citep{Baraffe}, the scatter into this bin from
brighter bins caused by photometric errors would not be modelled
correctly. We also excluded the two faintest bins from the probability
analysis as they are described as being incomplete (however we did
simulate them to correctly model scatter). Finally after going through
the probability analysis we excluded a secondary peak which appeared
at the edge of our grid as it was at a very high value of $\alpha$ excluded by
other studies. 
  \begin{figure}
 \setlength{\unitlength}{1mm}
 \begin{picture}(75,100)
 \includegraphics{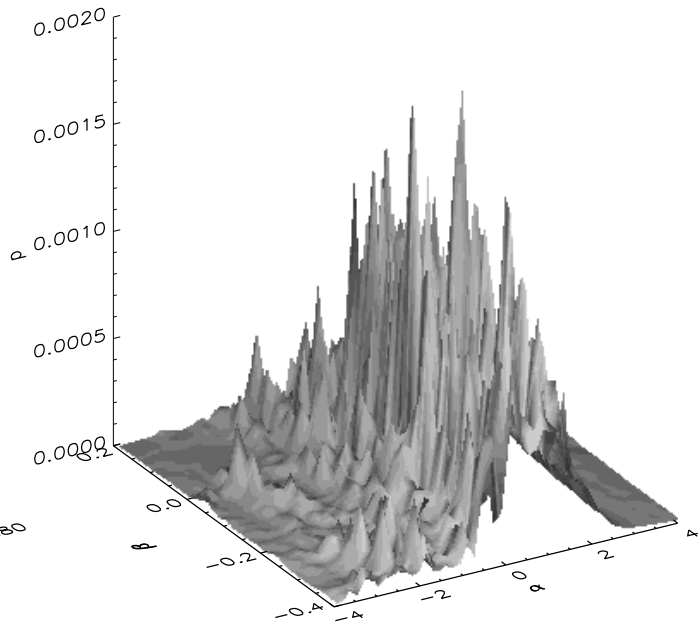}
 \end{picture}
 \caption[]{The probability surface produced using the \citet{Cruz2003}
 ultra--cool dwarf LF. The best fit parameters here are $\alpha=0.95
 \pm 1.17$ and $\beta=-0.134 \pm 0.173$.}
 \label{cruzfig}
 \end{figure}
The resulting probability surface is shown in
Figure~\ref{cruzfig}. The measured values of the parameters were;
$\alpha=0.95 \pm 1.17$ (implying a Mass Function rising at lower
masses) and $\beta=-0.134 \pm 0.173$. The question
remains over what range of masses is this result valid. If we take
$0.1M_\odot$ as a maximum mass then our minimum mass will by given by
the mass of an object with $M_J=14.0$ at an age of $10$~Gyr, the
maximum calculated in the \citet{Baraffe} models -- we find this to be
$\rm m=0.072M_\odot$. Hence the value of $\alpha$ covers the
range $0.072M_\odot< \rm m <0.1M_\odot$. This result (albeit with a large error)
is consistent with both \citet{Kroupa}, who measured a value in the range
$0.08M_\odot< \rm m <0.5M_\odot$ of $\alpha=0.3\pm0.5$, and with the
\citet{Chabrier2001} log--normal IMF peaking at $0.75M_\odot$. It differs
by just over $1\sigma$ from the \citet{Allen2005} value of
$-0.7\pm0.6$. The value of $\beta$ is consistent with a constant birthrate.

\section{Discussion}

The series of simulations presented here will have two main uses. In
the short term they can be used as a method for predicting the results
of the UKIDSS surveys, yielding more accurate values for the
expected number of extremely cool objects. This shows that -- given
current models and reasonable assumptions of the IMF and birthrate --
tens of Y dwarfs should be detected. The second use will be
using LAS data in conjunction with these simulations to constrain
underlying distributions. Our simulations show that for a sample
size with a typical local density in the range
$0.09M_\odot<m<0.1M_\odot$ the exponent of a power law IMF
can be constrained with an error of approximately $0.06$ while the
birthrate parameter $\beta$ can be constrained to an error of $0.016$.  

This method could prove very useful in determining
 both the IMF and birthrate. However binarity has
 not been taken into account in these simulations. The
 level of unresolved binarity will
 provide another parameter to characterise the results of the LAS.
Furthermore, the  
simulations do not take contamination of the sample into
account. Photometric errors will scatter objects such as hotter stars
and white dwarfs across colour--colour diagrams so that they have
colours similar to ultra--cool dwarfs. This contamination will have to
be quantified in order for accurate comparisons to be made with the
simulations. The quasar locus crosses the ultra--cool dwarf locus on a
$(J-H)$ versus $(H-K)$ plot 
(e.g.~\citealt{Leggett}), and such objects may also cause
contamination of the ultra--cool dwarf sample. 
However \citet{Hewett} have produced a method to separate 
quasars from ultra--cool
dwarfs using SDSS~\citep{Sloan} photometry, and this should
minimize quasar contamination.   

 \section{Conclusions}
 
Clearly
the techniques outlined here provide useful tools for both
 predicting the results of UKIDSS surveys and using those results to
 constrain the IMF (to an error in $\alpha$ of 0.06) and birthrate (to
 an error in $\beta$ of 0.015). 
We have demonstrated that using an existing small dataset (55
objects) we can utilise this technique to produce loose
constraints of $\alpha$ and $\beta$ that are consistent with other studies:   
 using the LF
 of \citet{Cruz2003}, we have found values of $\alpha=0.95 \pm 1.17$ in the
 range $0.072M_\odot< \rm m <0.1M_\odot$ and
 $\beta=-0.134 \pm 0.173$. These techniques are also
 complimentary to those used by \citet{Pinfield} to examine the IMF by empirical spectroscopic methods. While our techniques
 have the discussed limitations they will provide useful information
 to constrain the IMF and birthrates.  

\section*{Acknowledgments}

The authors would like to thank Sandy Leggett and Isabelle Baraffe for
providing unpublished data, and
Steve Warren, David Bacon, Thomas Kitching and Andy Taylor for  
helpful discussions. Thanks are also due to Sandy Leggett for refereeing
this paper and for making many useful suggestions that have resulted in 
a great improvement over the original manuscript.

\label{lastpage}

\end{document}